\documentclass[conference]{IEEEtran}
\IEEEoverridecommandlockouts
\usepackage{cite}
\usepackage{amsmath,amssymb,amsfonts}
\usepackage{algorithmic}
\usepackage{graphicx}
\usepackage{textcomp}
\usepackage{float}
\usepackage{tabularx}
\usepackage{adjustbox}
\usepackage{xcolor}
\usepackage{graphicx}
\usepackage{float}
\usepackage{multirow}
\def\BibTeX{{\rm B\kern-.05em{\sc i\kern-.025em b}\kern-.08em
    T\kern-.1667em\lower.7ex\hbox{E}\kern-.125emX}}
\begin{document}

\title{Enhancing Trustworthiness and Minimising Bias Issues in Leveraging Social Media Data for Disaster Management Response\\

}

\author{\IEEEauthorblockN{1\textsuperscript{st} Samia Abid}
\IEEEauthorblockA{\textit{Data Science, AI \& Modelling Centre (DAIM) (of Aff.)} \\
\textit{University of Hull (of Aff.)}\\
Hull, United Kingdom \\
samia.abid-2022a@hull.ac.uk}
\\
\IEEEauthorblockN{3\textsuperscript{rd} Dhaval Thakker}
\IEEEauthorblockA{\textit{School of Computer Science (of Aff.)} \\
\textit{University of Hull (of Aff.)}\\
Hull, United Kingdom \\
D.Thakker@hull.ac.uk}

\and
\IEEEauthorblockN{2\textsuperscript{nd} Bhupesh Mishra}
\IEEEauthorblockA{\textit{DAIM (of Aff.)} \\
\textit{University of Hull (of Aff.)}\\
Hull, United Kingdom \\
bhupesh.mishra@hull.ac.uk}
\\

\IEEEauthorblockN{4\textsuperscript{th} Nishikant Mishra}
\IEEEauthorblockA{\textit{Hull University Business School (of Aff.)} \\
\textit{University of Hull (of Aff.)}\\
Hull, United Kingdom \\
Nishikant.Mishra@hull.ac.uk}
\and

}

\maketitle

\begin{abstract}
Disaster events often unfold rapidly, necessitating a swift and effective response. Developing plans of action, resource allocation, and resolution of help requests in disaster scenarios is a time-consuming and complex process since the disaster-relevant information is often uncertain. Leveraging real-time data can significantly deal with data uncertainty and enhance disaster response efforts. To deal with uncertainty in data in real-time, social media appeared as an alternative effective source of real-time data as there has been extensive use of social media during and after the disasters. However, it also brings forth challenges regarding trustworthiness and bias in these data. To fully leverage social media data for disaster management, it becomes crucial to mitigate biases that may arise due to specific disaster types or regional contexts. Additionally, the presence of misinformation within social media data raises concerns about the reliability of data sources, potentially impeding actionable insights and leading to improper resource utilization. To overcome these challenges, our research aimed to investigate how to ensure trustworthiness and address biases in social media data. We aim to investigate to identify the factors that can be used to enhance trustworthiness and minimize bias to make an efficient and scalable disaster management system utilizing real-time social media posts, identify disaster-related keywords, and assess the severity of the disaster. By doing so, the integration of real-time social data can improve the speed and accuracy of disaster management systems.
\end{abstract}


\section{Introduction}
Disasters, whether natural or human-induced, have profound impacts on society. Its consequences include the human toll, environmental degradation, economic loss, psychological disruption, and infrastructure damage \cite{1}. Unfortunate events like Hurricane Sandy (2012), the Nepal Earthquake (2015), Hurricane Harvey (2017), Cyclone Idai (2019), the ongoing COVID-19 pandemic \cite{2}, and the Palestine-Israel conflict, etc have resulted in the loss of millions of lives, disrupted the economies, and strained healthcare systems. In such uncertain scenarios, effective disaster management and decision-making systems are central to mitigating and minimizing the effects of future calamities. However, not all emergency stakeholders possess specialized expertise in emergencies \cite{3}. 
\par 
The rise of social networks plays an important role in bridging the gap between stakeholders and those with more informed decision-making capabilities. Among various communication mediums, social media platforms have proven to be effective in disseminating real-time situational awareness, and safety instructions and facilitating rapid response. It has been reported that 5.04 billion people worldwide now use social media, a significant increase from the 3.6 billion users recorded in 2020 \cite{4}. Due to the accessibility of platforms like Facebook, Twitter, and Instagram, social media have been recognized as a powerful data source for studying disastrous events over the last decade. 
\par 
Compared to traditional methods of data collection like cameras \cite{4a}, RFID readers \cite{4b}, and GPS information \cite{4c}, social media data has promising merits: 1) scrapping posts from social media is economical, 2) the availability of user-generated content that can only be acquired through traditional data sources and government and regulatory organizations 3) real-time disaster updates, and 4) swift data acquisition. A massive amount and variety of user-generated content is shared on social media. These platforms enable a human-centric approach where the public can share rich footprints of disastrous events that can be used to enhance the effectiveness of disaster management systems. 
\begin{table*}
\caption{Biases in Disaster-relevant social media datasets}
\label{table1}
\begin{center}
\begin{tabular}{|c|c|c|c|c|c|}
\hline
\textbf{Dataset Name} & \textbf{Use Case} & \textbf{Platform} & \textbf{Geographical Bias} & \textbf{Language Bias} & \textbf{Crisis Representation Bias} \\
\hline
CrisisLexT26 & Crisis informatics research & Twitter & No & No & No \\
\hline
Hurricane Harvey Tweets & Disaster response research & Twitter & Yes (United States) & No & Yes (Hurricane Harvey) \\
\hline
Indonesian Tsunami Data & Crisis informatics research & Social media platforms & Yes (Indonesia) & Yes (Indonesian) & No \\
\hline
Nepal Earthquake Dataset & Crisis informatics research & Social media platforms & Yes (Nepal) & Yes (Nepali) & Yes (Nepal Earthquake) \\
\hline
COVID-19 Twitter Dataset & Pandemic response research & Twitter & No & No & No \\
\hline
\end{tabular}
\end{center}
\end{table*}
\par 
With potential benefits, several challenges are associated with the vast unstructured data coming from indeterminant sources \cite{5}. First, determining the veracity and quality of data sources is a significant issue. A reliable disaster management system demands trust in data for responsible disaster management. Moreover, to fully leverage social media data for disaster management, it is essential to aim at the conscious elimination of biases that may arise due to specific disaster types or regional contexts. Otherwise, these biases will be encoded into the latent representation of the disaster management system which in turn will result in unfair responses. This research work is dedicated to mitigating biases and ensuring the trustworthiness and quality of social media data for a fair and responsible disaster management and response system. We investigate the approaches that can enhance the veracity and quality of social media data. The approaches explored in this work can ensure the integrity of a scalable disaster management system. 

\par
The rest of the paper is organized as follows: Section II highlights the significant biases present in disaster-related social media data and provides effective mitigation strategies to address them. Section III focuses on trustworthiness concerns and the corresponding methods to tackle those issues. While Section IV concludes the presented work. 

\begin{figure*}
    \includegraphics[width=0.8\linewidth]{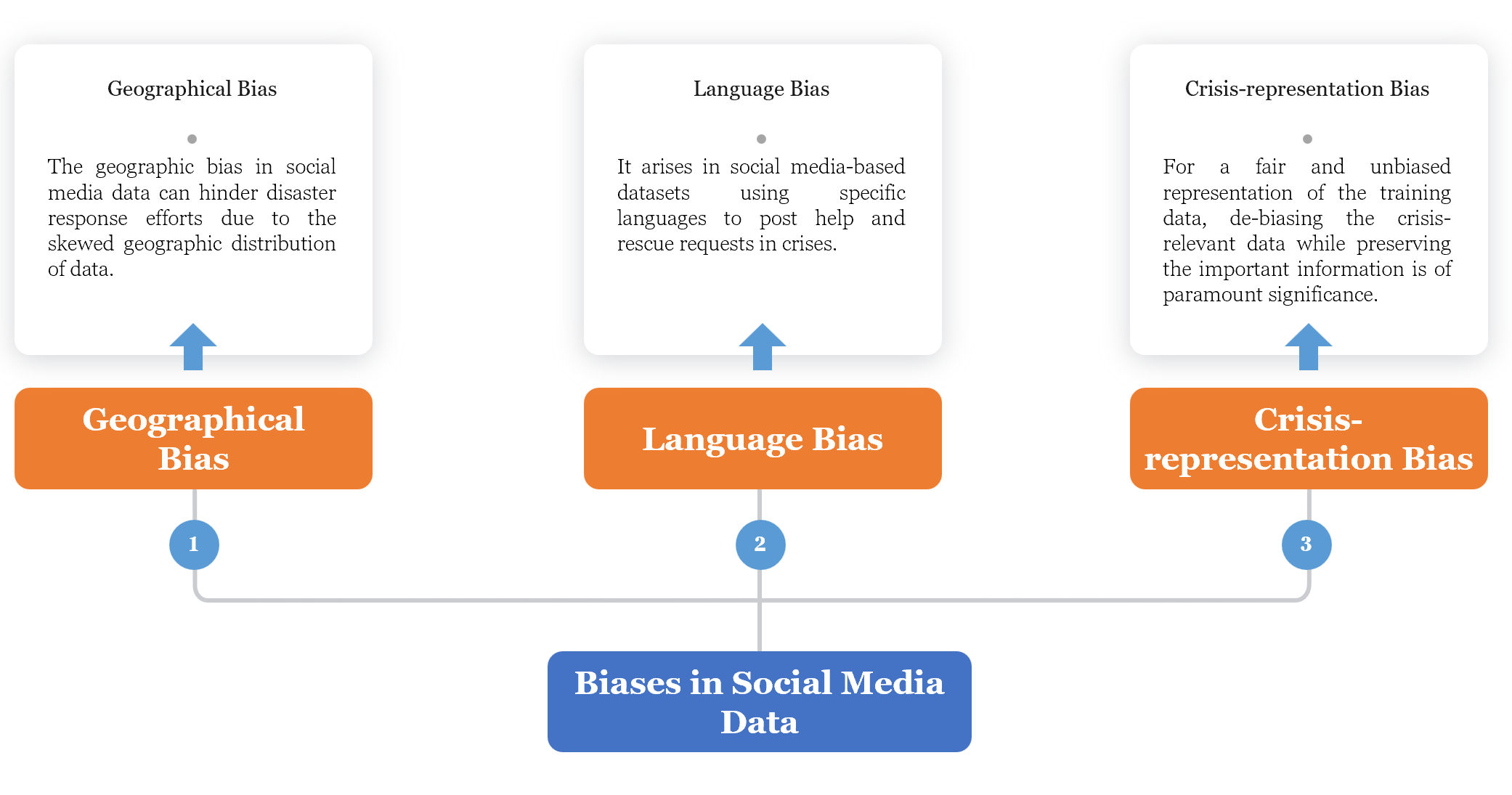}
    \centering
    \caption{Biases in Social Media Data for Disaster Management}
    \label{fig1}
\end{figure*}
\section{Biases in Social Media Data for Disaster Management}
Literature has witnessed the utilization of social media data in disaster management, contributing towards real-time crisis and public sentiments. Table \ref{table1} presents datasets used in the literature to advance research in disaster management and response.

\par 
The CrisisLexT26 dataset has been utilized for early detection of crisis-related events \cite{6}. CrisisNLP has been employed in research to develop a natural language processing tool to detect crisis-relevant information for humanitarian aid \cite{7}. Moreover, CrisisMMD has facilitated multimodal analysis of crises, exploring the integration of text, image, and video data for enhanced situational awareness \cite{8}. Similarly, the UnifiedCEHMET Dataset has demonstrated high precision in severity classification using deep learning methods \cite{9}. These studies exhibit the versatility and significance of these datasets in advancing understanding and improving response strategies in disaster management. However, the human-driven nature of these social media datasets can exhibit socio-demographics, language preferences, content, and spatial biases. The datasets contain inherent biases that hinder precise evaluation and response to crisis events, while the representation of disasters affects response strategies and resource allocation.
\\
Recently, this dilemma has caught the attention of researchers, and various biases have been identified. The identified biases are presented in \ref{fig1}. To mitigate their impact several approaches are  discussed in the subsequent subsections.
\subsection{Geographical Bias} 
The geographic bias in social media data can hinder disaster response efforts due to the skewed geographic distribution of data. Recent studies have highlighted that the uneven representation of certain locations or regions leads to disparities in data coverage during disaster situations \cite{10}. To address this issue, spatial sampling helps to enhance geographic representativeness by collecting data from diverse geographic regions. To quantify and visualize geographic biases in the dataset, sampbias has been developed as a tool to promote the usefulness of data in several research areas \cite{11}. Moreover, sampling techniques like the synthetic minority over-sampling technique (SMOTE) can improve the representativeness of the skewed instances from the majority regions \cite{12}.  By mitigating the geographic concentration of data, the machine-learning models will be more effective at capturing situations from social media in disasters.
\subsection{Language Bias} Language biases can limit the inclusivity and comprehensiveness of the data. It arises in social media-based datasets from the prevalent utilization of specific languages to post help and rescue requests in crises. According to \cite{13}, the language of posts shared on social media data is categorized into four types, including, i) global language, ii) local language, iii) mixed language, and iv) mixed script. Being the global language, most social media posts are shared in English. Potential countermeasures to demographic language bias are multilingual data collection \cite{14} and utilization of translation services.  The inception of large language models (LLMs) has advanced the field of natural language processing. Multilingual LLMs (MLLMs) address lingual biases because these language models are trained on multiple languages \cite{15}. Notable examples of MLLMs include mBERT \cite{16}, XLM-R \cite{17}, PaLM \cite{18}, mT5 \cite{19}, Falcon \cite{20}, BLOOM \cite{21}, and LLaMA \cite{22}. These MLLMs are trained on multiple languages and provide reasonably fair language processing capabilities. BLOOM is an outstanding MLLM due to its ability to support 46 languages, including French, English, African, Indonesian, Mandarin, etc. However, despite these advancements certain challenges like corpora, misalignment of MLLMs, and inherent biases in corpora still persist. To exemplify, ChatGPT is trained on 92.099\% of English Corpora and only 0.16\% of the corpora account for the Chinese language. True multilingualism can be achieved by creating a high-quality multilingual dataset. This area is undervalued and needs the attention of researchers.  
\subsection{Crisis-representation Bias} For a fair and unbiased representation of the training data, de-biasing the crisis-relevant data while preserving the important information is of paramount significance. To maintain the representation integrity, the learning fair representation (LFR) algorithm \cite{23} is designed to transform the training data while minimizing the loss of non-sensitive instances. Additionally, the prejudice-free representations (PFR) algorithm \cite{24} is proposed to identify and remove the features causing discrimination in the dataset, particularly those relevant to the sensitive attributes. By utilizing representation methods like LFR and PFR, the fairness of models in de-biasing disaster-relevant data can be ensured. 
\begin{figure}
    \includegraphics[width=1.0\linewidth]{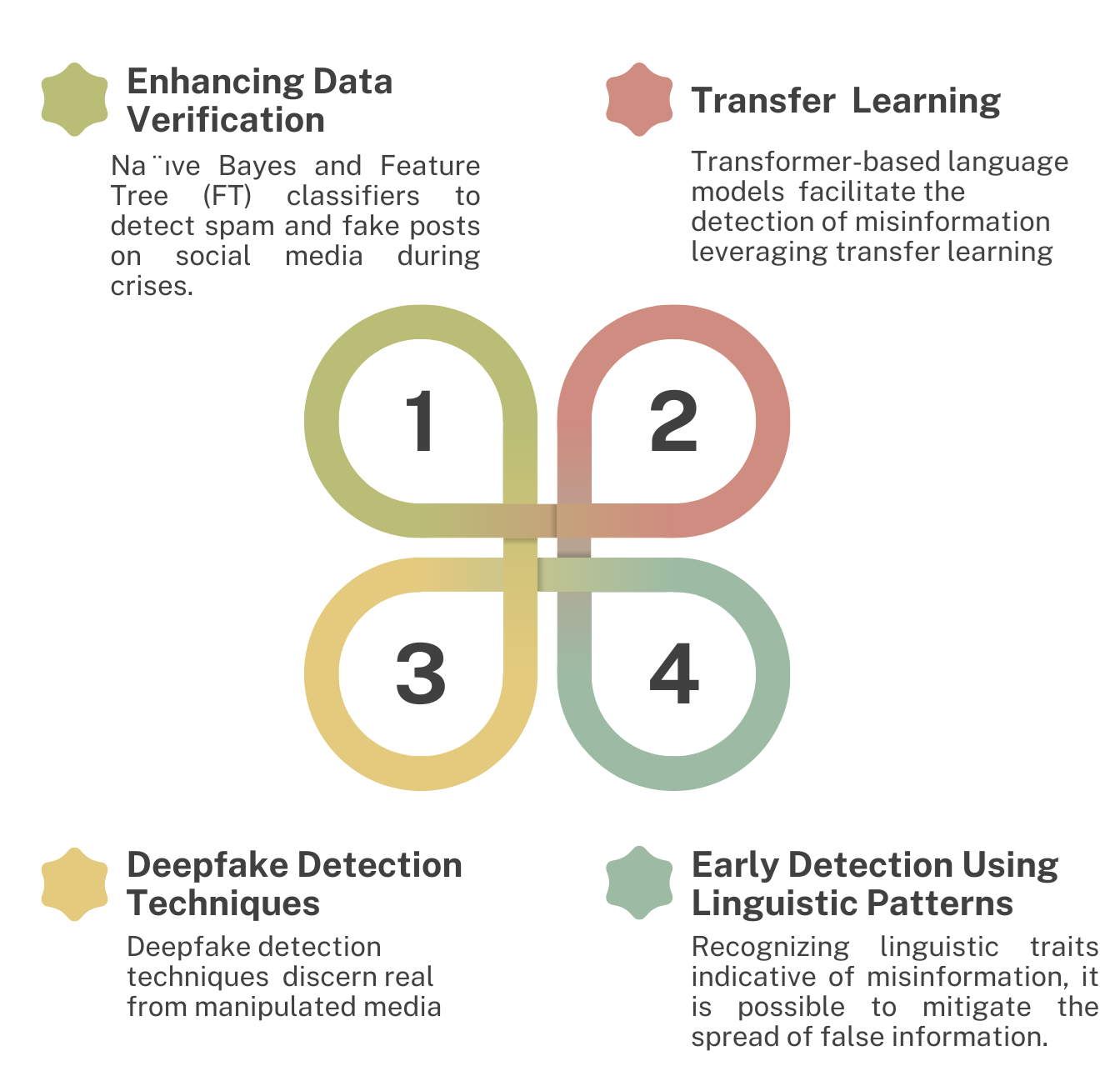}
    \caption{Techniques to Enhance Trustworthiness in Social Media Data}
    \label{fig2}
\end{figure}
\section{Trust in Disaster-relevant Social Media Data}
The trustworthiness of social media-based user-generated content is of paramount importance. The human-driven and diverse nature of the content and the spread of misinformation pose significant threats. It requires attention to assess the trustworthiness of the information provided on social media before using it for crisis-relevant applications, since the content may have been generated to spread false information. To combat these concerns and entrust social-media information relevant to sensitive events, various approaches have been proposed in the existing research (presented in Fig \ref{fig2}. In this section, we delve into existing techniques and mitigation strategies to ensure the quality and authenticity of social media data during disasters. 

\subsection{Social Media Users}
It is essential to understand the dynamics of disaster-relevant social media users that contribute to data quality and trustworthiness. The main actors that spread information in crises are non-governmental organizations, government agencies, research/academic bodies, and public profiles \cite{13}. Information disseminated through government organizations’ accounts is more trusted than that of a personal account. However, government authorities are noticed to be showing reduced participation in informing communities of the crises. Public users, on the other hand, have promising engagement on social media platforms. 
\subsection{Advanced Techniques to Enhance Trust in Social Media Data}
\begin{itemize}
\item Enhancing Data Verification: The verification of social media data emphasizes the development of mechanisms to assure trust. Algorithms such as Naïve Bayes and Feature Tree (FT) classifiers are used to detect spam and fake posts on social media during crises \cite{25}. These algorithms are utilized to distinguish between legitimate and false tweets with significance. 
Moreover, factors such as verified accounts, premium subscription services, and engagement metrics play an important role in ensuring trust in social media content and combatting misinformation propagation. In Twitter, factors like verified users, blue tick subscriptions, and retweet metrics enhance the trust level of their shared content. 
\item Transfer Learning: Advanced transformer-based language models like BERT or RoBERTa leveraging transfer learning facilitate the detection of misinformation conveyed on social media platforms \cite{26}. Fine-tuning these models on misinformation datasets can help the system identify misleading information. Thereby, the reliability of the information disseminated on social media platforms can be enhanced. 
\item Deepfake Detection Techniques: Researchers are exploring various approaches utilizing Deepfake detection techniques to discern real from manipulated media \cite{27}. Computer vision methodologies and deep neural networks are among the approaches that can be employed to mitigate potential implications leading to distrust.
\item Early Detection Using Linguistic Patterns: Early identification of fraudulent information on social media platforms is made possible by linguistic cue analysis linked to misinformation and user attribute inference based on linguistic traits. By understanding various patterns like how the information spreads and recognizing linguistic traits indicative of misinformation, it is possible to mitigate the spread of false information. For instance, authors in \cite{28} address the spread of misinformation on Twitter during the COVID-19 pandemic. The authors analyzed textual and non-textual cues to understand their influence on retweeting behavior and the spread of false information on social media. By analyzing 4923 tweets featuring disaster-relevant hashtags in May 2020, the study aims to discern retweet probability and volume. For this purpose, they have focused on employing logistic regression and machine learning techniques. 

\end{itemize}
These approaches can be leveraged in disaster management and response applications to enhance trust in social media data. These methods facilitate effective management for resource allocation and rescue during disastrous events.  

\section{Conclusion}

Disaster, either natural or human-induced, necessitates swift response to the situations for effective allocation of resources to mitigate the loss and save lives. The prevalent utilization of social media data has made it possible due to rich user-generated content which can help disaster management systems locate the victims and volunteers. However, social media data has challenges in terms of certain biases and trustworthiness. To encounter these issues, we have identified three types of biases that need to be mitigated before the utilization of social media data for sensitive crisis-relevant decision-making systems. We also highlight the importance of verifying the authenticity and quality of information posted on these platforms to ensure trustworthiness. For this purpose, we have explored several techniques from the literature to ensure trust in social media data by identifying misleading content. Incorporating the factors that can be used to enhance trustworthiness and minimize bias can help to develop an efficient and scalable disaster management system utilizing real-time social media posts. In the future, we intend to conduct rigorous analysis by considering case studies to provide empirical evidence. Furthermore, we will delve into the challenges and limitations of the mitigation strategies discussed in this research by focusing on concerns regarding implementation, resource constraints, and adaptability. It will help in the successful adoption of the proposed strategies and highlight the need for future research in this area.

\vspace{12pt}

\end{document}